\documentclass[prd, aps, superscriptaddress, preprintnumbers, twocolumn, floatfix, nofootinbib]{revtex4-2}
\pdfoutput=1

\usepackage{amsfonts}
\usepackage{amsmath}
\usepackage{amssymb}
\usepackage{bm}
\usepackage{dcolumn}
\usepackage{graphicx}   
\usepackage[latin1]{inputenc}
\usepackage{latexsym}
\usepackage{rotating}
\usepackage{hyperref}
\usepackage{graphicx}
\usepackage{color}
\usepackage{cases}

\newcommand\be{\begin{equation}}
\newcommand\ba{\begin{eqnarray}}
\newcommand\ee{\end{equation}}
\newcommand\ea{\end{eqnarray}}

\newcommand{\gsim}{\mathrel{\hbox{\rlap{\lower.55ex \hbox {$\sim$}}
                   \kern-.3em \raise.4ex \hbox{$>$}}}}
\newcommand{\lsim}{\mathrel{\hbox{\rlap{\lower.55ex \hbox {$\sim$}}
                   \kern-.3em \raise.4ex \hbox{$<$}}}}

\begin{document}

\title {N-Body Simulation of Early Structure Formation from Cosmic String Loops}

\author{Hao Jiao}
\email{hao.jiao@mail.mcgill.ca}
\affiliation{Department of Physics, McGill University, Montr\'{e}al, QC, H3A 2T8, Canada} 

\author{Robert Brandenberger}
\email{rhb@physics.mcgill.ca}
\affiliation{Department of Physics, McGill University, Montr\'{e}al, QC, H3A 2T8, Canada} 

\author{Alexandre Refregier}
\email{alexandre.refregier@phys.ethz.ch}
\affiliation{Institute for Particle Physics and Astrophysics, ETH, Wolfgang-Pauli-Strasse 27, CH-8093 Zurich, Switzerland}

\date{\today}


\begin{abstract}

By means of N-body simulations, we study early structure formation in the presence of a scaling distribution of cosmic string loops.  Cosmic string loops dominate the high redshift halo mass function while the fluctuations seeded by the standard structure formation scenario dominate structure at low redshifts. In our study, the effects of the cosmic string loops are taken into account by displacing the dark matter particles and their velocities at the initial time of the simulation by amounts determined by the analytical analysis which makes use of the Zeldovich approximation.  We find that the resulting halo mass function is to a good approximation given by the sum of the analytically determined cosmic string halo mass function and the halo mass function obtained from the standard $\Lambda$CDM model.

\end{abstract}

\pacs{98.80.Cq}
\maketitle

\section{Introduction} 
\label{sec:intro}

A subset of particle physics models beyond the Standard Model have solutions corresponding to cosmic string defects (see e.g. \cite{VS, HK, RHBrev} for reviews of cosmic strings and their role in early universe cosmology). If Nature is described by such a model, then causality arguments \cite{Kibble} imply that a network of cosmic strings will form in the early universe and persist to the present time.  Cosmic strings correspond to lines of trapped energy,  and the induced gravitational effects lead to signatures in cosmology.  The gravitational effects of strings depend on only one free parameter, namely the string tension $\mu$ which is of the order $\eta^2$, where $\eta$ is the energy scale of the phase transition which leads to defect formation.  Searching for the signals of cosmic strings in the sky is hence a way to probe particle physics ``from top down'' (since the effects are larger for larger values of $\eta$), while usual accelerator searches probe new physics ``from bottom up'' (since they are more sensitive to physics at lower values of $\eta$).

After the phase transition during which strings form, the distribution of strings rapidly approaches a ``scaling solution'' \cite{Kibble} according to which the statistical properties of the string distribution are invariant in time $t$ if all lengths are scaled to the Hubble radius $t$.\footnote{In this work, we use the natural units with $c=k_B=\hbar=1$, so the Hubble radius is $r_H = H^{-1}\sim t$.} The string network consists of a random-walk-like network of ``long strings'' (strings with curvature radius comparable to or larger than $t$) and a distribution of string loops which result from the inter-commutation of long strings.  While the scaling distribution of long strings is robust since it is derived from general causality arguments, the distribution of string loops is less certain since it depends on the decay channels of string loops. It is generally believed that gravitational radiation \cite{VV} dominates the loop decay, but some field theory simulations indicate that particle emission might have a large effect \cite{Hind}.  In this paper we shall work in the context of the ``one-scale model'' of the distribution of string loops \cite{onescale} which is supported by the numerical simulations of \cite{CSsimuls}. According to this model, at any given time $t$ there are string loops with radii $R$ in the range $\gamma G\mu t < R < \alpha t$, where the constant $\alpha$ indicates the mean loop radius $R = \alpha t$ at the time $t$ when the loop is formed, and $\gamma \sim 10^2$ is a constant determined by the strength of gravitational radiation from loop oscillations.  The number density in comoving coordinates $n(R, t) dR$ of loops in the radius interval between $R$ and $R + dR$ is (for times after the time of matter and radiation equality, $t_{eq}$) is given by
\be
n(R,t) \, = \, N\alpha^2\beta^{-2}t_0^{-2}R^{-2} \label{eq-nLoop-1}
\ee
for $\alpha t_{eq} \leq R \leq \alpha t$, and
\be
n(R, t) \, = \, N\alpha^{5/2}\beta^{-5/2}t_{eq}^{1/2}t_0^{-2}R^{-5/2} \label{eq-nLoop-2}
\ee
for $\gamma G\mu t < R < \alpha t_{eq}$,  where $N$ is a constant determined by the number of long strings per Hubble volume, and $\beta R$ is the mean length of a loop with radius $R$. Loops with radius smaller than $\gamma G\mu t$ live for less than a Hubble expansion time and their comoving number density can be taken to be independent of $R$.

The strongest robust bound on the string tension stems from the angular power spectrum of cosmic microwave background (CMB) anisotropies and is $G\mu < 10^{-7}$ \cite{Dvorkin}. By looking at signals of long strings in angular maps and making use of dedicated statistics such as wavelets the bound could be strengthened \cite{Hergt}.  Long strings also yield distinctive signals in 21-cm redshift maps \cite{Holder}, and 21-cm surveys which probe the epoch before reionization have the potential to yield comparable bounds \cite{Maibach}.

\begin{figure}[h!]
  \includegraphics[width=8cm]{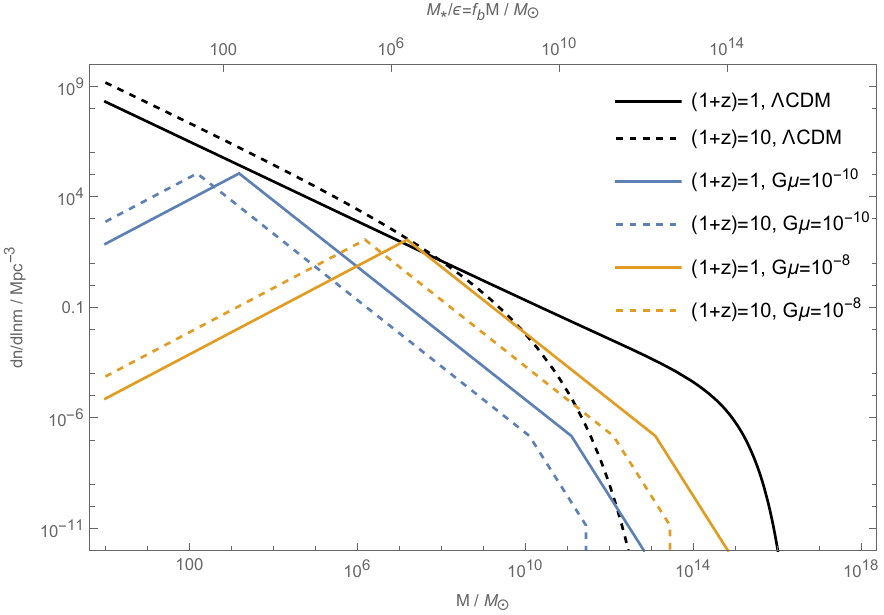}
\caption{Comparison of the halo mass function sourced by a scaling distribution of string loops with the corresponding mass function in the $\Lambda CDM$ model.  The halo mass is indicated on the bottom horizontal axis (while the top one indicates the stellar mass as a function of the stellar formation efficiency parameter $\epsilon$ and the baryon fraction $f_b$). The black curves show the halo mass function in the $\Lambda CDM$ model for redshifts $z+1 = 1$ (solid curves) and $z+1 = 10$ (dashed curves). In orange and blue are the corresponding halo mass functions due to strings at the corresponding redshifts.  The results for two interesting values of the string tension are shown: $G\mu = 10^{-8}$ in orange and $G\mu = 10^{-10}$ in blue.   While the $\Lambda CDM$ mass function dominates at late times,  the effect of the string loops is more important at higher redshifts. The turnover redshift above which the strings dominate depends on $G\mu$.  Considering redshifts close to the turnover, we see that the string loops dominate at the higher mass end while the $\Lambda CDM$ fluctuations are more important for smaller masses. This can be seen by considering the mass functions at redshift $z = 10$ for $G\mu = 10^{-8}$.}
\label{Fig1}
\end{figure}

Since strings form nonlinear density fluctuations beginning at the time when they form, string loops will dominate early nonlinear structure formation and could explain the origin of high redshift super-massive black holes \cite{SMBH} (see e.g. \cite{Marta} for a review on super-massive black holes). In a previous paper \cite{us}, we have computed the halo mass function obtained from a scaling distribution of string loops.  As expected, we found that the resulting halo mass function dominates over the corresponding mass function from the standard $\Lambda CDM$ scenario (based on Gaussian adiabatic primordial fluctuations) at high redshifts. This is illustrated in Figure~\ref{Fig1} which is an updated version \footnote{Compared to the work of \cite{us}, we here included an extra factor of $2/5$ in the accretion by a string loop of mass, following the prescription in \cite{VS}.} of the corresponding figure in \cite{us}.  Our previous work \cite{us} indicates that for a string tension in the range $10^{-9} < G\mu < 10^{-8}$, cosmic strings could explain recent JWST results \cite{JWST} which indicate an over-abundance of high redshift galaxies compared to what the standard $\Lambda CDM$ model predicts.

In our previous work, we did not consider the interplay between fluctuations seeded by strings and those generated by the $\Lambda CDM$ perturbations. At high redshifts, this is a reasonable approximation, but as soon as the $\Lambda CDM$-induced inhomogeneities become important, this approximation breaks down, and the interplay between the fluctuations seeded by the two sources must be considered. Here, we take a first step towards addressing this challenge.  We perform N-body simulations of structure formation with both of the effects of string loops and $\Lambda CDM$ fluctuations included.  The bottom line of our study is that the resulting halo mass function closely follows that mass function obtained by adding the mass functions of the individual sources. This result will be useful in analyzing the effects of cosmic string loops in the mildly nonlinear phase of structure formation.

\section{Numerical Setup} \label{review}

Our numerical simulations make use of the Gadget-2 code \cite{Gadget} with $\Lambda CDM$ initial conditions generated from Planck-normalized Gaussian $\Lambda CDM$ fluctuations by means of the N-GencIC code \cite{N-Gen}.\footnote{We use the python package LensTools \cite{LensTools} to read and write Gadget snapshots.}  The initial conditions for the N-body simulations were set at a redshift of $z + 1 = 32$, deep in the region where linear perturbation theory can be trusted.  The resulting initial positions and velocities of the particles were then displaced according to the effects which a scaling distribution of cosmic string loops would have, computed in the Zeldovich approximation \cite{Zeld}.  This procedure is discussed in detail in the Appendix.

The Rockstar code \cite{Rockstar} is used to identify halos in the output of the simulations.  We ran simulations with $N_{particle} = 256^3$ particles in a box of comoving box size $100 h^{-1} {\rm{Mpc}}$, where $h$ is the value of the Hubble constant in units of $100 {\rm{km}}/{\rm{sec Mpc}}$ and we set it to be $0.7$ in our simulations.

The following figures show snapshots of our results. Figures~\ref{Fig2} and \ref{Fig3} show the distribution of the mass points in a pure $\Lambda CDM$ simulation at a redshifts of $z + 1 = 4$ and $z + 1 = 1$, respectively, while Figures~\ref{Fig4} and \ref{Fig5} are corresponding to snapshots at the same redshifts if the effects of a scaling distribution of cosmic strings with a large value $G\mu = 10^{-7}$ of the string tension are added. Note that the Gaussian seeds were taken to be the same in the simulations with and without the strings.  As expected, the effects of cosmic strings are very difficult to be identified by eye at these low redshifts. Only a couple of cosmic string loop-seeded halos can be seen at redshift $z+1 = 4$.

\begin{figure}[h!]
  \includegraphics[width=10cm]{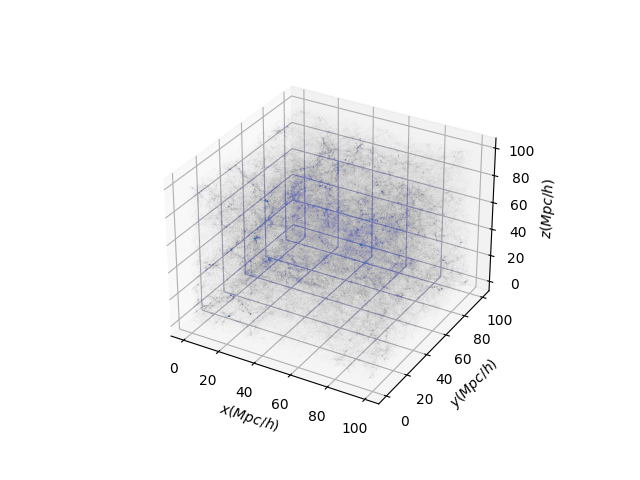}
\caption{Distribution of dark matter particles in a pure $\Lambda CDM$ simulation at redshift $z + 1 = 4$. These particles are illustrated by semi-transparent, pale blue points, and overdensities are shown by a deeper shade of blue. }
\label{Fig2}
\end{figure}

\begin{figure}[h!]
  \includegraphics[width=10cm]{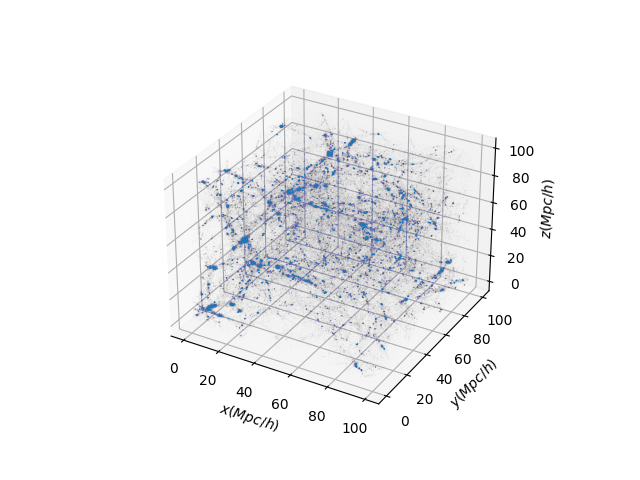}
\caption{The distribution of dark matter particles in the same simulation as Fig. 2 at redshift $z + 1 = 1$.}
\label{Fig3}
\end{figure}

\begin{figure}[h!]
  \includegraphics[width=10cm]{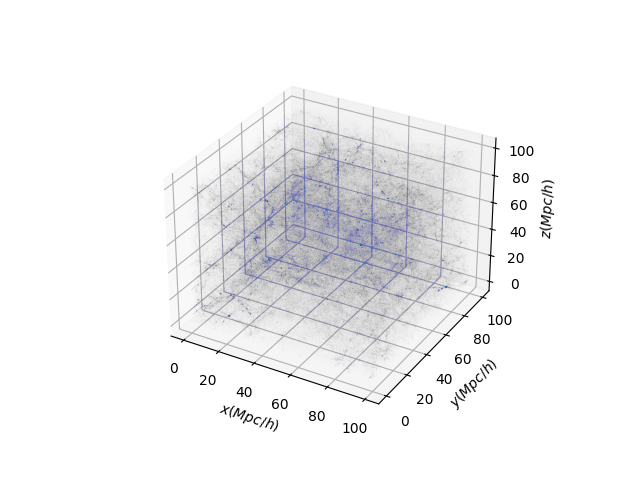}
\caption{Distribution of dark matter particles in a simulation which includes both $\Lambda CDM$ fluctuations (with the same seeds as those in Fig. 2) and cosmic strings with tension $G\mu = 10^{-7}$ at redshift $z + 1 = 4$.}
\label{Fig4}
\end{figure}

\begin{figure}[h!]
  \includegraphics[width=10cm]{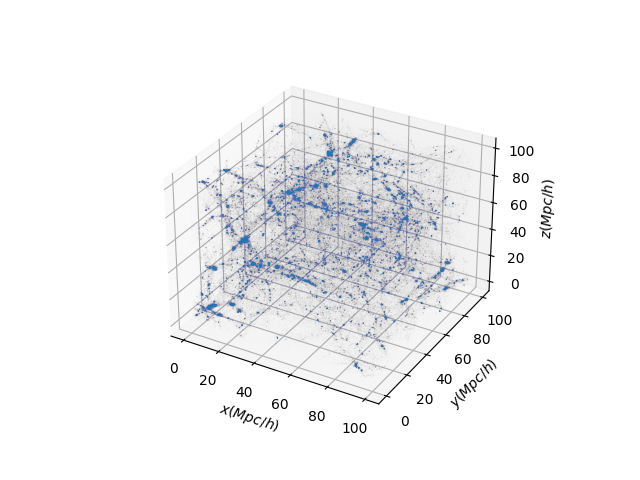}
\caption{Distribution of dark matter particles in the simulation of Fig. 4 at redshift $z + 1 = 1$.}
\label{Fig5}
\end{figure}

\begin{figure}[h!]
  \includegraphics[width=8cm]{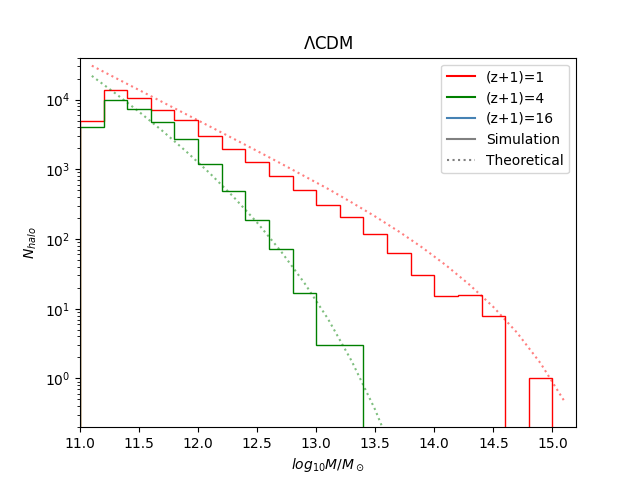}
\caption{Comparison between the numerically obtained halo mass function (solid curves) with the analytical approximation (dotted curves) in a pure $\Lambda CDM$ simulation at two redshifts.  The vertical axis gives the number of halos in the mass bin with width $d\lg (M/M_{\odot}) = 0.2$ in the simulation volume.  Note that there are too few halos at a redshift $z + 1 = 16$ to be found in the simulation volume, and are therefore not shown.}
\label{Fig6}
\end{figure}

The code was tested in various ways.  First,  the numerically obtained halo mass function in a pure $\Lambda CDM$ model was compared to the analytical results obtained using the Press-Schechter \cite{PS} formalism.  The numerical and analytical results are indeed in good agreement as is shown in Figure~\ref{Fig6}. As is apparent, there are no nonlinear halos at high redshifts.

\begin{figure}[h!]
\includegraphics[width=8cm]{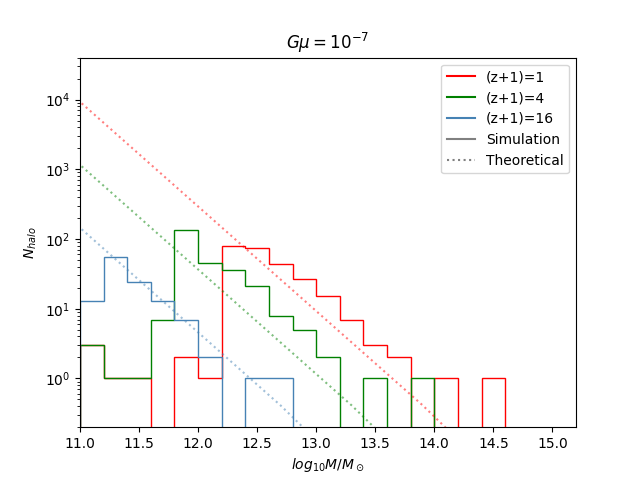}
\caption{Comparison between the numerically obtained halo mass function (solid curves) with the analytical approximation (dotted curves) in a pure cosmic string simulation with $G\mu = 10^{-7}$. The results at three redshifts are shown. Note that cosmic strings lead to large mass nonlinear halos at high redshifts while the standard $\Lambda CDM$ model does not.}
\label{Fig7}
\end{figure}

Figure~\ref{Fig7} shows the corresponding mass functions in a simulation with only cosmic strings (with tension $G\mu = 10^{-7}$). The numerical results are the solid curves, and the analytical curves are given by the dotted lines.  First of all, we note the good agreement between the numerical and analytical curves for large masses.  Secondly, we note that cosmic strings seed large mass nonlinear halos at high redshifts. 

Now that we have hopefully persuaded the reader that our code is working correctly, we can turn to the results.

\section{Results} 
  
\begin{figure}[h!]
\includegraphics[width=8cm]{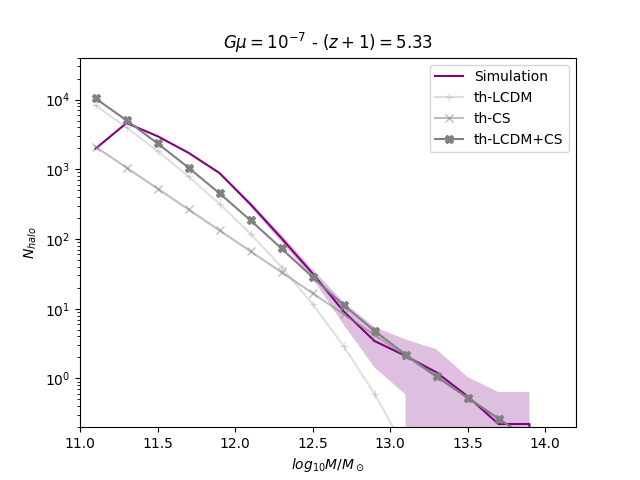}
\caption{Mass function in simulations with both cosmic strings ($G\mu = 10^{-7}$) and $\Lambda CDM$ fluctuations at a redshift $1 + z = 5.33$ when the theoretical cosmic string and $\Lambda CDM$ mass functions cross. The numerical simulations show that the resulting mass function closely follows the sum of the two theory curves. }
\label{Fig8}
\end{figure}

\begin{figure}[h!]
\includegraphics[width=8cm]{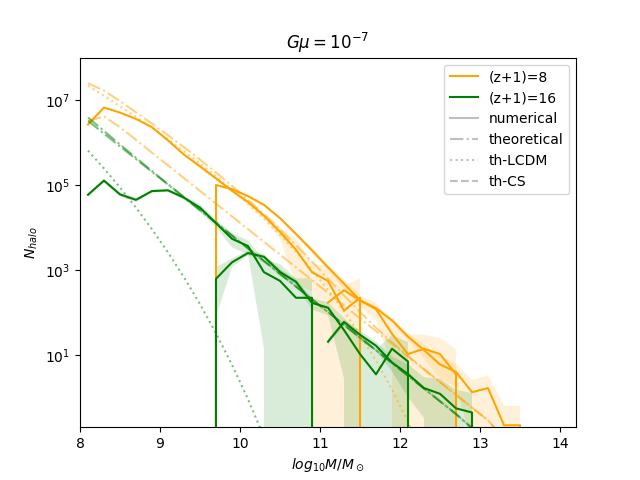}
\caption{Mass function in a simulation with both cosmic strings ($G\mu = 10^{-7}$) and $\Lambda CDM$ fluctuations at redshifts  $z + 1 = 8$ (yellow) and $z + 1 = 16$ (green).  Here, simulations in three different box sizes were patched together, and thus the envelope of the three numerical curves for each color should be considered.  The dotted curves correspond to the $\Lambda CDM$ theory predictions, the dashed one to the cosmic string predictions, and the dot-dashed curve to the total theory curves (note that these faint curves are also color-coded).  The numerically determined halo mass function follows the total theory curve except for small masses, which is due to the mass resolution of these simulations.}
\label{Fig9}
\end{figure}

\begin{figure}[h!]
  \includegraphics[width=8cm]{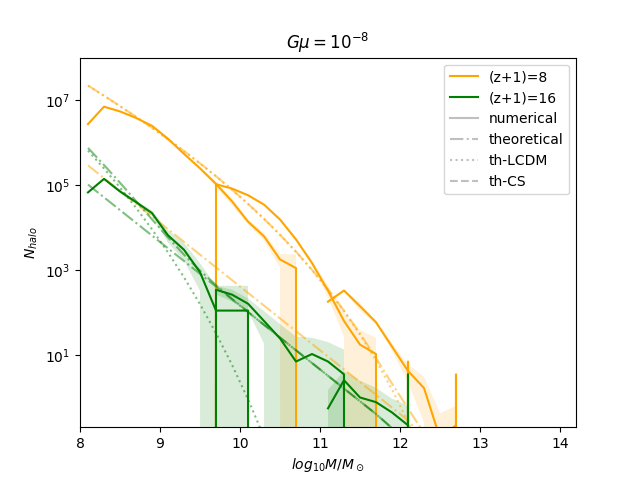}
\caption{Same as Figure 9 but for $G\mu = 10^{-8}$.}
\label{Fig10}
\end{figure}

We have performed a series of N-body simulations \footnote{We ran 9 simulations with different N-GenIC seeds and different distributions of loops following the number density in \eqref{eq-nLoop-1} and \eqref{eq-nLoop-2} and random spatial distribution to get the expectation and r.m.s. (shaded region in Figures~\ref{Fig8}-\ref{Fig10}) of the halo mass function.} containing both $\Lambda CDM$ fluctuations and a scaling distribution of string loops with various values of the string tension $G\mu$. The key question we wish to address is how the numerically computed halo mass function compares to the sum of the analytically computed string halo mass function superposed with the analytical $\Lambda CDM$ mass function obtained by means of the Press-Schechter model.
 
Figure~\ref{Fig8} shows the results for simulations with $G\mu = 10^{-7}$ at a redshift $z + 1 = 5.33$, a redshift chosen because for the value of $G\mu$ used, the two analytical mass functions cross - the string-induced mass function (indicated by th-CS) dominating for larger masses while the $\Lambda CDM$ curve (indicated by th-LCDM) is higher at the low mass end.  The results of the numerical analysis are shown in color (with the spread indicating the variance of the results). We see that the results agree well with the theory curve (indicated by th-LCDM+CS) which is the addition of the cosmic string and $\Lambda CDM$ theory curves.

With the small number of mass points which we have simulated we have a limited mass resolution. To obtain results for a wider range of masses we have patched together simulations (including both strings and $\Lambda CDM$ fluctuations) in different sized boxes\footnote{Here we consider three different box sizes - $10^2h^{-1}$Mpc, $10^{3/2}h^{-1}$Mpc, and $10h^{-1}$Mpc, and run 9 simulations with $N_{particle}= 256^3$ particles for each of them.}.  The resulting mass functions are shown in Figures~\ref{Fig9} and \ref{Fig10}, the former for $G\mu = 10^{-7}$ and the latter for $G\mu = 10^{-8}$.

\section{Discussion}

We have studied early nonlinear structure formation for a model in which, in addition to the standard Gaussian $\Lambda CDM$ fluctuations,  there is a scaling distribution of cosmic string loops.  We find that the resulting numerically computed halo mass function matches well a theoretical curve which is the superposition of the string loop-induced halo mass function and the usual $\Lambda CDM$ halo mass function.  This result will allow us to more reliably compute the effects of cosmic strings on early structure formation in the mildly nonlinear region, in particular in the redshift range of reionization. 

Our analysis demonstrates that $\Lambda CDM$ fluctuations do not interfere with the role that cosmic string loops could play in early structure formation,  e.g. in generating the seeds for high redshift super-massive black holes \cite{SMBH} and high redshift galaxies \cite{us}.  The string tensions we have explored are below the current robust upper bound, but higher than the value of $G\mu \sim 10^{-10.5}$ which is preferred if cosmic strings are to explain the recently detected \cite{PTA} gravitational wave signal in millisecond pulsar timing arrays (PTAs) \cite{PTA-CS} (see also \cite{CS-pulsar} for earlier work). In this context, it is important to take into account that it is smaller loops which dominate the PTA signal while it is larger loops which are relevant for super-massive black hole formation, and for explaining early galaxy formation.  The small loop distribution is more affected by the unknown physics that goes into establishing a formula for the loop distribution function such as the ``one-scale model'' than the distribution of larger loops. Hence,  cosmic strings with tension $G\mu > 10^{-10.5}$ could well be consistent with the PTA constraints.

\section*{Acknowledgement}

\noindent RB wishes to thank the Pauli Center and the Institutes of Theoretical Physics and of Particle- and Astrophysics of the ETH for hospitality. The research of RB at McGill is supported in part by funds from NSERC and from the Canada Research Chair program.  
HJ wishes to thank Juan Gallego,  Matteo Puel,  Pascale Berner,  Pascal Hitz and Uwe Schmitt for help with the numerics.

\section*{Appendix: Accretion onto a String Loop}

Here we explain how we modify the initial conditions for our Gadget-2 simulations to take into account the effect of a scaling distribution of string loops.  We start with a distribution of dark matter particles generated from Planck-normalized Gaussian $\Lambda$CDM fluctuations obtained from the N-GencIC code.  We then consider a distribution of cosmic string loops with a size distribution obtained from the cosmic string scaling solution described in the Introduction section, and with uncorrelated loop positions.  

We use the Zel'dovich approximation \cite{Zeld} to calculate the displacement and velocity of dark matter particles due to a loop which are used in the initial conditions for the Gadget-2 simulation.  For each dark matter particle in the simulation, we consider the change in position and velocity induced by string loops, treating the effects of each loop independently. 

The Zel'dovich approximation considers the evolution of mass shells surrounding the string loops. The Newtonian gravity effect of the loop causes the Hubble expansion of the shell to slow down, and eventually the shell ``turns around'', i.e. the physical distance of the shell from the loops stops increasing. The Zeldovich approximation is only valid before the particles turn around.  After turnaround, we assume that the particles virialize before the initial time of the numerical simulation.

It is justified to view the cosmic string loop as a point source since the separation of the particles in our simulation is large compared to the loop size (note that the impact of the finite size of oscillating loops is studied in \cite{JBB}). Therefore, the accretion onto a loop is spherically symmetric and thus we only need to calculate the radial component of the coordinates and velocities of the particles relative to the center of the loop.

In the Zel'dovich approximation, we consider the physical distance $h(q, t)$ of a shell with initial comoving distance $q$ from the loop, and denote by $\psi(q, t)$ the comoving displacement of the shell as a consequence of the gravitational attraction. We consider only accretion in the matter-dominated era and hence take the scale factor to be $a(t) = (t / t_0)^{2/3}$. The relation between $h$ and $\psi$ is
\be
h(q, t) \, = \, a(t) \bigl( q  - \psi(q, t) \bigr) \, . 
\ee
The dynamics of the test particle is described by Newtonian gravity, i.e.
\be
\ddot h \, = \, -\frac{\partial \Phi}{\partial h} \, ,
\ee
where the gravitational potential $\Phi$ is determined by the Poisson equation
\be
\nabla^2\Phi  \, = \, 4\pi G(\rho_{bg}+\rho_{\rm{string}}) \, ,
\ee
where $\rho_{\rm{string}}=M_{\rm{loop}}\delta(\mathbf{x})$ is the energy density due to the string loop (taken to be at the origin of the coordinate system). These two equations can be combined to yield the following equation of motion for $\psi$:
\be
\ddot\psi+\frac{4}{3}t^{-1} \dot\psi-\frac23 t^{-2}\psi \, = \, \frac{GM_{\rm{loop}}}{q^2}\left(\frac{t_0}{t}\right)^2,
\ee
where $M_{\rm{loop}} = \beta \mu R$ is the total mass of the string loop with radius $R$. With initial conditions $\psi(t_i) = {\dot{\psi}}(t_i) = 0$ the solution is
\be \label{distance}
\psi(q,t) \, = \, \frac{3}{2}\frac{GM_{loop}t_0^2}{q^2}\left[-1+\frac{3}{5}\bigg(\frac{t}{t_i}\bigg)^{2/3}+\frac{2}{5}\bigg(\frac{t_i}{t}\bigg)\right],
\ee
Here,  $t_i$ is the time that the loop begins to accrete, which is $t_{eq}$ for loops formed in the radiation phase, while for loop formed in the matter phase we have $t_i = \alpha R / \beta$, which is the formation time of this loop.  The comoving velocity of the test particles is
\be \label{velocity}
\dot\psi(t) \, = \, \frac{3}{5}\frac{GM_{loop}\,t_0^2}{q^2\,t_i}\left[\bigg(\frac{t_i}{t}\bigg)^{1/3}-\bigg(\frac{t_i}{t}\bigg)^{2}\right].
\ee

We can calculate the radius of the turnaround shell by solving the equation
\be
\dot h (q_{nl}(t),t) \, = \, 0 \, ,
\ee
which yields
\be
q_{nl} \,  \simeq \, \bigg(\frac{9}{5}GMt_0^2\bigg)^{1/3}\bigg(\frac{t}{t_i}\bigg)^{2/9} \, .
\ee 
For particles with distance $q > q_{nl}(t)$ (where $t$ is taken to be the initial time of the N-body simulation) from a loop, we model the effect of the string loop by adding displacements and velocities towards the loop given by (\ref{distance}) and (\ref{velocity}) to their original coordinates and velocities.

For particles with distance $q < q_{nl}(t)$ from one of the loops, we assume that the particle has virialized in the halo created by the loop. This implies that the physical distance of the particle from the loop will be half of the physical turnaround radius.

The physical distance of the particle at the time $t_{ta}$ of turnaround is
\be
h(t_{ta}) \, = \, a(t_{ta})(q-\psi(t_{ta})) = \frac12 a(t_{ta})q \, .
\ee
Thus, the physical and comoving distances of the particles from the loop will be given by
\ba
h_{vir}(q) \, &=& \, \frac12 h(t_{ta},q) \, = \, \frac14 a(t_{ta})q,\\
r^c_{vir}(q, z) \, &=&  \frac{1}{a(t)}h_{vir}(q) \, = \, \frac{1}{4} \frac{a(t_{ta})}{a(t)}q \, = \, \frac{1}{4}\frac{z}{z_{ta}}q \, ,
\ea
where $z_{ta}$ is the redshift corresponding to the turnaround time $t_{ta}$. This redshift can be determined by solving ${\dot{h}}(q, t_{ta}(q)) = 0$ and yields
\be
\left(\frac{z_i}{z_{ta}}\right) \, \simeq \, \frac{5q^3}{9GM_{loop}t_0^2} \, .
\ee
Thus, we obtain
\be
r_{vir}^c(z)\equiv q-\psi(z) \, \simeq \, \frac{5}{36}  \frac{q^4}{GM_{loop}t_0^2}\frac{z}{z_i}
\ee

The velocity of test particles can be computed by requiring the physical height of the particle to be constant. This leads to a comoving velocity of a virialized particle towards the string loop of magnitude
\be
\dot h_{vir} \, = \, \dot a r_{vir}^{c}+a\dot r_{vir}^c = 0
\ee
which implies
\be
\dot\psi \, = \, \dot r_{vir}^c  \, = \, H r_{vir}^c \, =\,  \frac{2}{3t}r_{vir}^c \, .
\ee

Note that for dark matter particles which are within the virialized radius of a particular loop,  no effects from other loops are taken into account.


\end{document}